\documentclass[aps,prl,
reprint,superscriptaddress,longbibliography]{revtex4-1}
\usepackage{amsmath,amssymb,mathtools,bm}
\usepackage{tikz,newtxtext,newtxmath}
\usetikzlibrary{arrows.meta,calc}
\usepackage{graphicx}
\usepackage{hyperref}
\hypersetup{hidelinks}
\usepackage{booktabs}
\newcommand{\ERW}{\mathrm{ERW}}
\newcommand{\WTD}{\mathrm{WTD}}
\newcommand{\dd}{\mathrm{d}}
\newcommand{\Bset}{\mathcal{B}}
\newcommand{\Hset}{\mathcal{H}}
\newcommand{\Lset}{\mathcal{L}}
\newcommand{\Uset}{\mathcal{U}}
\newcommand{\Vset}{\mathcal{V}}
\begin{document}

\title{Excursion-Resolved Thermodynamic Inference without Observing Reverse Transitions}
\author{Jie Gu}
\email{jiegu1989@gmail.com}
\affiliation{Chengdu Academy of Educational Sciences, Chengdu 610036, China}
\date{\today}

\begin{abstract}
Experiments often detect only selected transitions, leaving the underlying state dynamics and dissipation hidden. We show that waiting-time distributions between directed visible events can reconstruct the boundary dynamics without observing the reverse of every detected transition. Under a simple source-target closure condition, they determine visible rates and the hidden excursions connecting observed states, yielding a rigorous lower bound on total entropy production that improves the standard waiting-time estimate when reverse events are available. Remarkably, observing only a spanning directed cycle is sufficient to identify the full Markov generator, even when all reverse cycle edges and other microscopic transitions remain unseen.
\end{abstract}

\maketitle

\paragraph{Introduction.---} A molecular-motor step, an ion-channel opening, or a tunneling event is often much easier to detect than the microscopic state trajectory that produced it.
Transition-level measurements are therefore natural in single-molecule experiments and mesoscopic transport, whereas dissipation is generated by complete stochastic paths through states and cycles that may remain unresolved
\cite{Sekimoto2010,Jarzynski2011,Seifert2012,VanDenBroeck2015,Peliti2021,Schmiedl2007,Ge2012,Rao2016}.
This mismatch has motivated thermodynamic inference from currents, fluctuations, time asymmetry, and partial trajectory statistics
\cite{Barato2015,Gingrich2016,Horowitz2020,Li2019,Vu2020,Dechant2021,Kawai2007,Parrondo2009,Roldan2010,Roldan2012} (see Refs. \cite{Seifert2019,Seifert2026Review,GhosalBisker2026Irreversibility,Dieball2025,AlemanyRibezziCrivellariRitort2015,GodecMakarov2023Challenges} for reviews).
It has also made the choice of experimentally accessible coarse variables part of the thermodynamic inference problem itself
\cite{Rahav2007,Esposito2012,Bo2014,Bo2017,Seiferth2020,Teza2020,Hartich2021,Hartich2023,Zhao2024}.

A particularly direct protocol records only selected jumps and the times elapsed between successive recorded events.
Waiting-time statistics can reveal irreversibility, constrain hidden cycles, diagnose hidden disorder, and expose microscopic paths and topology
\cite{Skinner2021PRL,Skinner2021PNAS,Ehrich2021,Harunari2022,VanDerMeer2022,VanDerMeer2023,Maier2024Periodic,Maier2025}.
Harunari \textit{et al.} showed that occurrence and intertransition-time statistics of selected transitions carry information about dissipation and hidden dynamics \cite{Harunari2022}, while Van der Meer \textit{et al.} formulated reverse-resolved transition waiting times as a semi-Markov process whose physical time reversal yields an entropy-production estimator \cite{VanDerMeer2022}.
More recent work has used the short-time structure of the same statistics to infer hidden paths and kinetic topology \cite{Maier2024Periodic,Maier2025}.
These developments leave two related questions open.
First, existing thermodynamic constructions naturally compare observed transitions with their observed reverse partners. How much of the hidden dynamics and its irreversibility can be reconstructed from a genuinely directed event record when those reverse transitions are not measured?
Second, recent counterexamples show that the standard WTD bound can be weaker than a thermodynamic-uncertainty bound constructed from the current of the same partially observed link \cite{Gu2026}.
This suggests that the conventional WTD estimator does not exhaust the thermodynamic information available in the event-time statistics, raising a complementary question of whether the same waiting-time record can be exploited more fully to yield a tighter lower bound on entropy production.

The key observation is that pairwise visible reversibility is stronger than what reconstruction actually requires.
Let $\Lset\neq\varnothing$ denote the observed directed transitions, and let $\Uset\equiv\{u_\ell:\ell\in\Lset\}$ and $\Vset \equiv \{v_\ell:\ell\in\Lset\}$ be the sets of their microscopic source and target states.
We require only the source--target closure condition
\begin{equation}
\Uset = \Vset.
\label{eq:closure}
\end{equation}
Define the boundary $\Bset=\Uset=\Vset$ under this condition. Pairwise visible reversibility becomes a special case.

\begin{figure*}[t]
\centering
\includegraphics[width=\textwidth]{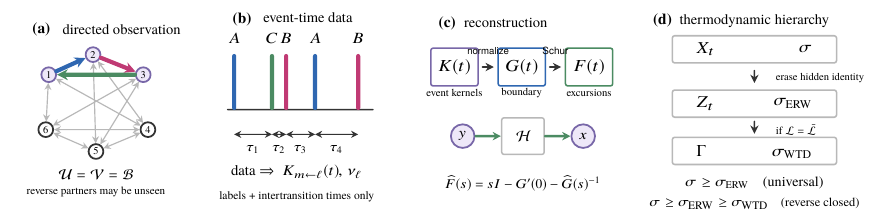}
\caption{From directed transition records to reconstructed dissipation.
(a) A source-target-closed set of one-way visible events can be embedded in a network whose reverse transitions and internal motion remain unseen.
(b) The measured record contains only visible labels and intertransition times, from which the complete pairwise kernel matrix $K(t)=[K_{m\leftarrow\ell}(t)]_{m,\ell}$ and event rates $\nu_\ell$ are estimated.
(c) Zero-time normalization reconstructs the boundary propagator $G(t)$ from $K(t)$, while a Schur complement reconstructs the hidden-excursion kernel $F(t)$.
(d) The boundary-path law reconstructed for $Z_t$ retains more time-reversal information than the visible event path $\Gamma$: universally $\sigma\geq\sigma_{\ERW}$, and for reverse-closed visible sets $\sigma\geq\sigma_{\ERW}\geq\sigma_{\WTD}$.}
\label{fig:physical}
\end{figure*}

In this Letter, we show that this broader observation geometry is sufficient to reconstruct substantially more than the visible event process itself.
The path-space irreversibility defines an excursion-resolved waiting-time estimator $\sigma_{\ERW}$ satisfying the universal bound $\sigma\geq\sigma_{\ERW}$.
When reverse partners are also observed, the same raw event-time record yields the conventional waiting-time estimator and the hierarchy sharpens to
$\sigma\geq\sigma_{\ERW}\geq\sigma_{\WTD}$.
Thus ERW identifies thermodynamic information that is present in the cross-waiting-time laws but discarded by the standard event-level description.
Finally, directed detector cycles provide a simple experimental realization of source--target closure: extending the cycle over nested boundary sets cannot decrease the resolved dissipation. More generally, any source--target-closed detector set whose boundary spans the full microscopic state space identifies the full generator and hence the full entropy-production rate; a spanning one-way cycle is a particularly sparse realization, requiring neither reverse cycle edges nor any noncycle transitions to be observed.
Fig.~\ref{fig:physical} summarizes this reconstruction and information hierarchy.

\paragraph{Directed events as boundary probes.---}
Consider an irreducible stationary continuous-time Markov jump process $(X_t)_{t\geq0}$ on a finite state space $\Omega$, with column generator $W$, where $W_{ij}$ is the rate for $j\to i$.
States are even under time reversal, rates are time independent, and each ordered state pair denotes a single transition channel.
The detector identifies known source and target states, resolves every occurrence of its selected transitions, and measures time in a calibrated physical unit.
We assume microscopic dynamical reversibility, $W_{ij}>0$ whenever $W_{ji}>0$, while the detector set itself need not be reverse closed.
A visible event $\ell$ is the directed jump $u_\ell\to v_\ell$ with microscopic rate $k_\ell=W_{v_\ell u_\ell}$.
The stationary record contains only the event labels $\ell_n$ and the intertransition times $\tau_n$.
Its complete two-event statistics are the joint next-label/time densities $K_{m\leftarrow\ell}(t)$, collected into the matrix $K$, where $K_{m\leftarrow\ell}(t)\dd t$ is the conditional probability that visible event $m$ occurs next after a delay in $[t,t+\dd t]$ given that event $\ell$ has just occurred.
Thus $\sum_m\int_0^\infty K_{m\leftarrow\ell}(t)\dd t=1$. An individual kernel is not separately normalized.

Delete every visible off-diagonal rate from $W$ while retaining the original diagonal escape rates, and call the resulting killed generator $S$.
Immediately after event $\ell$, the microscopic state is known to be $v_\ell$.
For event $m$ to occur next after a delay $t$, the killed process must survive without any intervening visible event, propagate from $v_\ell$ to the source state $u_m$, and then make the visible jump $m$.
The standard first-passage construction therefore gives
\cite{Harunari2022,VanDerMeer2022}
\begin{equation}
K_{m\leftarrow\ell}(t)
=
k_m[e^{St}]_{u_m v_\ell}.
\label{eq:kernelprop}
\end{equation}

The short-time limit of transition waiting-time distributions has previously been used to recover the rate of an observable transition when its reverse transition is also observed
\cite{Maier2024Periodic, MaierHasler2026}.
Equation~\eqref{eq:kernelprop} shows that the reverse transition itself is not required.
Under the source--target closure condition in Eq.~\eqref{eq:closure}, choose any visible event $\ell_x$ ending at $x$ and any visible event $m_x$ starting at $x$.
Then
\begin{equation}
k_m
=
K_{m\leftarrow\ell_{u_m}}(0^+),
\qquad
G_{xy}(t)
\equiv
[e^{St}]_{xy}
=
\frac{K_{m_x\leftarrow\ell_y}(t)}{k_{m_x}}.
\label{eq:reconstruct}
\end{equation}
Here $K(0^+)$ denotes the right-hand short-time limit of the waiting-time density, not a finite probability of a zero-duration waiting interval.
The first identity extends the reverse-pair reconstruction of Ref.~\cite{Maier2024Periodic}: any detected event ending at the source of $m$ prepares the required microscopic state.
The second identity follows directly from the first-passage representation in Eq.~\eqref{eq:kernelprop}.
Let $P_{\Bset}$ denote the coordinate projector onto the boundary states. Hence the same directed event record determines every visible rate and the full boundary block $G(t)=P_{\Bset}e^{St}P_{\Bset}$ without requiring every detected transition to be accompanied by its reverse.
If several visible events share a source or target, Eq.~\eqref{eq:reconstruct} gives redundant estimates of the same rate or propagator element, providing internal consistency checks.


\paragraph{Hidden motion as a measurable excursion kernel.---}
Partition the killed generator into boundary and hidden states as $S=\bigl(\begin{smallmatrix}A&U\\V&Q\end{smallmatrix}\bigr)$, where $\Hset$ contains all states outside $\Bset$.
A path can leave a boundary state $y$, wander through $\Hset$, and re-enter the boundary at $x$ after duration $t$.
The rate density of such a hidden excursion is the matrix
\begin{equation}
F(t)=Ue^{Qt}V,
\qquad
\widehat F(s)=sI_{|\Bset|}-G'(0)-\widehat G(s)^{-1},
\label{eq:schur}
\end{equation}
where $\widehat M(s)\equiv\int_0^\infty e^{-st}M(t)\,\dd t$ for any matrix-valued function $M(t)$. 
Detailed derivations are provided in the End Matter.
The second identity is the Schur complement of the hidden block, valid for real $s\geq0$ in this finite irreducible setting.
Here $F(t)$ has units of inverse time squared: $F_{xy}(t)\dd t$ is an excursion-initiation rate resolved by its subsequent duration, not a normalized first-passage density.
It shows that the entire all-time excursion kernel is fixed by the same pairwise waiting-time matrix without identifying the number, topology, or individual rates of hidden states.
This complements the hidden-path program of Ref.~\cite{Maier2025}, where the first nonzero short-time coefficients identify shortest compatible paths and constrain their microscopic entropy production.
The short-time expansion resolves which paths appear first, while Eq.~\eqref{eq:schur} resums every compatible hidden path into the return dynamics seen by the observable boundary.
Related first-passage, memory-kernel, and coarse-graining constructions appear throughout stochastic processes and nonequilibrium thermodynamics \cite{Gillespie1977,Pigolotti2008,Puglisi2010,Knoch2015,WangQian2007,Bo2017,Maes2009,Andrieux2008}.

\paragraph{Excursion-resolved entropy production.---}
The boundary stationary probabilities are observable under source-target closure.
For any visible event $m$ leaving $x$, its stationary occurrence rate obeys $\nu_m=p_x^{\rm ss}k_m$, which gives $p_x^{\rm ss}=\nu_m/k_m$.
These are absolute stationary probabilities and must not be renormalized to sum to one on $\Bset$ when hidden states remain.
The direct microscopic rates between boundary states are also fixed because $G'(0)=S_{\Bset\Bset}$ contains every unobserved boundary jump and only lacks the visible directed rates removed from $S$.
Explicitly, for $x\neq y$, the direct boundary-to-boundary rate $\kappa_{xy}\equiv W_{xy}=G'_{xy}(0)+\sum_{\ell\in\Lset:\,u_\ell=y,\,v_\ell=x}k_\ell$, as derived  in Eq.~\eqref{eq:restore}.
Define $J^{\rm d}_{xy}=p_y^{\rm ss}\kappa_{xy}$ and the excursion flux density $J^{\rm e}_{xy}(t)=p_y^{\rm ss}F_{xy}(t)$.
The irreversibility of the boundary-resolved trajectory is
\begin{equation}
\begin{aligned}
\sigma_{\ERW}={}&
\sum_{x<y\in\Bset}
\left(J^{\rm d}_{xy}-J^{\rm d}_{yx}\right)
\ln\frac{J^{\rm d}_{xy}}{J^{\rm d}_{yx}}
\\
&+\sum_{x<y\in\Bset}\int_0^\infty
\left[J^{\rm e}_{xy}(t)-J^{\rm e}_{yx}(t)\right]
\ln\frac{J^{\rm e}_{xy}(t)}{J^{\rm e}_{yx}(t)}\dd t.
\end{aligned}
\label{eq:erw}
\end{equation}
Every term compares a physical boundary process with its time reverse.
The first line resolves direct boundary jumps.
The second resolves hidden excursions by entrance state, exit state, and duration.
An excursion that leaves $y$ and returns to $y$ is self-reversed at this coarse level and contributes no entropy, while its unresolved internal cycles can still contribute to the gap $\sigma-\sigma_{\ERW}$.

The bound follows from a physical coarse trajectory rather than from an auxiliary dynamics.
Define $Z_t=x$ whenever the microscopic state equals $x\in\Bset$ and $Z_t=H$ for all states in $\Hset$.
This pointwise map commutes with physical time reversal.
The boundary holding rates, direct jumps, and excursion kernels determine the stationary law of $Z$, not the unobserved realization of $Z$ in an individual measured interval. Its path-space relative-entropy rate is Eq.~\eqref{eq:erw}, and data processing gives the general directed-observation result \cite{CoverThomas2006,Kawai2007,GomezMarin2008,Parrondo2009,Shiraishi2015,Polettini2017,Bisker2017,Uhl2018}
\begin{equation}
\sigma\geq\sigma_{\ERW}.
\label{eq:directedbound}
\end{equation}

\begin{figure}[t]
\centering
\includegraphics[width=0.8\columnwidth]{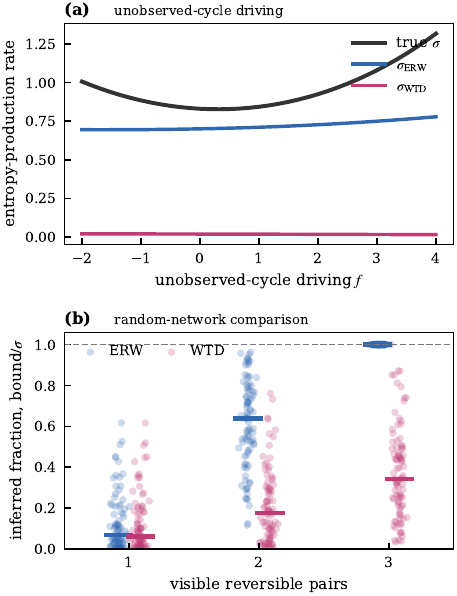}
\caption{Reverse-closed comparison of ERW and WTD from the same event-time record.
(a) An unobserved-cycle driving scan for a six-state network with two visible reversible edge pairs ($|\Omega|=6$ and $|\Bset|=4$). The scanned links are disjoint from the detected pairs.
The exact entropy production, the excursion-resolved bound $\sigma_{\ERW}$, and the standard transition-WTD bound $\sigma_{\WTD}$ are shown together.
(b) Ninety random six-state networks are analyzed for observations with one, two, or three visible reversible edge pairs.
Points show the inferred fractions $\sigma_{\ERW}/\sigma$ and $\sigma_{\WTD}/\sigma$, and horizontal bars show medians.
ERW systematically sharpens the same-record WTD estimate while remaining below the exact entropy production. See the End Matter for numerical details.}
\label{fig:reconstruction}
\end{figure}

\paragraph{Reverse-closed observations recover the WTD hierarchy.---}
For a visible event $\ell:u_\ell\to v_\ell$, let $\bar\ell$ denote the reversed transition $v_\ell\to u_\ell$, and define $\bar{\Lset}\equiv\{\bar\ell:\ell\in\Lset\}$. Suppose now that the visible set is additionally closed under reversal, $\Lset=\bar{\Lset}$.
The event-time trajectory then has a physical reverse that can be represented using the same detected transitions, and the standard transition-WTD estimator can be written as \cite{Harunari2022,VanDerMeer2022}
\begin{equation}
\sigma_{\WTD}=
\sum_{\ell,m}\nu_\ell\int_0^\infty
K_{m\leftarrow\ell}(t)
\ln\frac{K_{m\leftarrow\ell}(t)}{K_{\bar\ell\leftarrow\bar m}(t)}\dd t.
\label{eq:wtd}
\end{equation}
Erasing from $Z_t$ every boundary motion except the selected visible transitions produces this event path.
A second data-processing step therefore gives
\begin{equation}
\sigma\geq\sigma_{\ERW}\geq\sigma_{\WTD}.
\label{eq:hierarchy}
\end{equation}
This comparison uses exactly the same raw list of visible labels and event times for ERW and WTD.
Both estimators are functionals of the same pairwise event-time kernels $K_{m\leftarrow\ell}(t)$, but ERW asks more of their detailed shape: it uses zero-time limits to identify visible rates, differentiates the reconstructed propagator, and inverts its Laplace-domain matrix through Eq.~\eqref{eq:schur}, and recovers $F(t)$ by an inverse Laplace transform.
The additional cost is therefore inferential rather than experimental.
Finite records can make this distinction important because short-time limits, derivatives, and matrix inversions generally require more regularization than evaluating the WTD likelihood-ratio functional in Eq.~\eqref{eq:wtd} \cite{Fritz2025,Ertel2022}.

Figure~\ref{fig:reconstruction} compares the excursion-resolved bound with the standard WTD estimator for a measurement setting in which every detected transition is accompanied by its reverse, so that both estimators are computed from the same event-time record.
Panel~(a) uses a six-state network with detected pairs $1\leftrightarrow2$ and $3\leftrightarrow4$ and varies the affinity of the entirely unobserved cycle $1\to5\to6\to1$, changing the exact entropy production and the ERW and WTD bounds while leaving the four detected microscopic rates fixed.
Panel~(b) repeats the comparison for ninety random six-state networks while varying the number of visible reversible edge pairs from one to three.
Throughout the scan and the random ensemble one finds $\sigma_{\WTD}\leq\sigma_{\ERW}\leq\sigma$, and the gain of ERW over WTD typically grows as more reverse-closed detector pairs are included.

The physical content of the strict improvement is also transparent.
Standard WTD compresses all boundary-resolved motion inside one observed waiting interval into an initial event, a final event, and a total duration.
ERW resolves which boundary states are visited before that compression.
The log-sum inequality is strict on an event-time fiber when its conditional forward-to-reverse likelihood ratio is nonconstant on a set of positive conditional measure. A strictly positive improvement of entropy-production rates requires this information loss to be extensive in time, not merely an endpoint effect.
This generalizes the repeated-versus-alternated distinction emphasized by Harunari \textit{et al.} \cite{Harunari2022}.
A transition pair can be thermodynamically silent in Eq.~\eqref{eq:wtd} and still determine an element of $G(t)$ that changes the excursion kernel and the ERW bound.

\begin{figure}[t]
\centering
\includegraphics[width=\columnwidth]{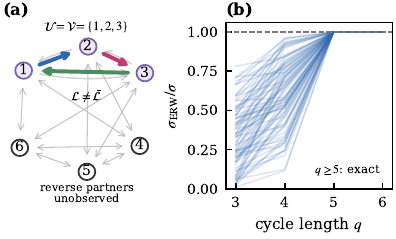}
\caption{Thermodynamic inference without visible reversibility.
(a) A one-way detector cycle provides source-target closure on the boundary states $\{1,2,3\}$ while remaining non-reverse-closed, $\Lset\neq\bar\Lset$.
The reverse partners exist microscopically but are not detected.
(b) Ninety random six-state networks are analyzed using directed cycles of length $q=3,4,5,6$.
Each faint polyline shows $\sigma_{\ERW}/\sigma$ for one realization as the cycle length increases.
The recovered fraction is nondecreasing along the nested boundaries $\Bset_q=\{1,\ldots,q\}$. For these six-state networks, $q=5$ and $q=6$ are exact because at most one hidden state remains.}
\label{fig:directed}
\end{figure}

\paragraph{Source-target-closed detector geometries.---}
Source-target closure makes the extension beyond visible reversibility experimentally concrete.
A detector set containing only $1\to2$, $2\to3$, and $3\to1$ already supplies an incoming and an outgoing anchor at each boundary state.
Its reverse transitions can remain completely unseen.
Figure~\ref{fig:directed} illustrates this one-way detector geometry and shows how the recoverable dissipation grows as the directed cycle is extended to cover more states.
No standard same-record WTD entropy estimator of the form Eq.~\eqref{eq:wtd} is available when only this one-way set of transitions is detected.

The spanning-cycle limit is a useful identifiability corollary rather than the central mechanism.
If $\Bset$ equals the full microscopic state space, there is no hidden block and $G(t)=e^{St}$ is the complete killed propagator.
The measured zero-time kernels give every visible $k_\ell$, so
\begin{equation}
\begin{aligned}
&W=G'(0)+\sum_{\ell\in\Lset}k_\ell|v_\ell\rangle\langle u_\ell|,\\
&\sigma_{\ERW} =\sigma, \qquad (\Bset = \Omega).
\end{aligned}
\label{eq:fullreconstruction}
\end{equation}
A spanning directed cycle therefore suffices even if none of its reverse partners is detected and every noncycle transition is experimentally invisible.
The nontrivial information comes from cross-waiting-time laws rather than from continuous state observation.
Each detected jump prepares a known endpoint, and every other detector probes how the killed dynamics propagates probability from that endpoint to its own source state.
The full WTD matrix is therefore a transfer matrix measurement of the hidden network.
If exactly one hidden state remains, the symbol $H$ uniquely identifies that state, so $X\mapsto Z$ is still one-to-one and $\sigma_{\ERW}=\sigma$.
This explains the $q=5$ saturation in Fig.~\ref{fig:directed}. 

This viewpoint suggests a detector-design principle that is more general than reversible edge coverage.
A useful set of transition detectors should create boundary states that are both arrival and departure anchors.
If a larger experimental record does not satisfy Eq.~\eqref{eq:closure}, any source-target-closed subset can still be analyzed by retaining only those event labels and recomputing the waiting times between them.
Directed cycles are especially simple closed subsets, while more general observation graphs need not consist of reverse pairs or even be strongly connected.
Nested boundary sets give a monotone hierarchy because the coarser boundary path is obtained by relabeling newly resolved states as hidden,
\begin{equation}
\sigma_{\ERW}(\Bset_1)\leq\sigma_{\ERW}(\Bset_2)\leq\sigma, \quad (\Bset_1\subseteq\Bset_2).
\label{eq:boundaryhierarchy}
\end{equation}
The value of adding a detector is therefore tied to the dynamical landmarks it creates and to the cross kernels it makes accessible.

\paragraph{Where the missing dissipation resides.---}
For finite entropy-production rates, the hierarchy admits an exact information-theoretic decomposition.
Because $Z$ is a deterministic function of the microscopic path, the chain rule separates the irreversibility retained by $Z$ from microscopic information lost within each coarse-path fiber.
Let $P_X^T$ and $P_X^{R,T}$ denote the forward and physically time-reversed microscopic path measures on $[0,T]$, and let $P_Z^T$ and $P_Z^{R,T}$ be their pushforwards under $X\mapsto Z$.
In the stationary long-time limit,
\begin{equation}
\begin{aligned}
\sigma-\sigma_{\ERW}
={}&\lim_{T\to\infty}\frac{1}{T}\mathbb E_{z\sim P_Z^T}\\
&\quad D_{\rm KL}\!\left(P_X^T(\cdot\mid z)\Vert P_X^{R,T}(\cdot\mid z)\right)
\geq0.
\end{aligned}
\label{eq:uppergap}
\end{equation}
Thus the upper gap is the conditional irreversibility of microscopic paths consistent with a fixed boundary path.
It vanishes when the microscopic forward-to-reverse likelihood ratio is determined by $Z$. The spanning-cycle construction in Eq.~\eqref{eq:fullreconstruction} is sufficient but not necessary.
Hidden realizations may therefore share the same $G(t)$ and $F(t)$ yet differ through cycles internal to $\Hset$.

When every detected transition is accompanied by its reverse, the contraction $Z\mapsto\Gamma$ similarly gives, with $P_\Gamma^T$ and $P_\Gamma^{R,T}$ denoting the forward and reversed visible event-path measures,
\begin{equation}
\begin{aligned}
\sigma_{\ERW}-\sigma_{\WTD}
={}&\lim_{T\to\infty}\frac{1}{T}\mathbb E_{\gamma\sim P_\Gamma^T}\\
&\quad D_{\rm KL}\!\left(P_Z^T(\cdot\mid\gamma)\Vert P_Z^{R,T}(\cdot\mid\gamma)\right)
\geq0.
\end{aligned}
\label{eq:lowergap}
\end{equation}
Hence $\sigma_{\ERW}-\sigma_{\WTD}$ is the conditional irreversibility erased when a boundary-resolved path is reduced to visible labels and waiting times.
Equality holds when the forward-to-reverse likelihood ratio of $Z$ is fixed by $\Gamma$. Endpoint-only differences cannot produce a positive rate gap.
Additional boundary visits within a detected waiting interval provide the simplest source of a strict gap, and even a kernel with no direct contribution to Eq.~\eqref{eq:wtd} can increase ERW by changing these conditional boundary paths.

In excursion form, $\ln[J^{\rm e}_{xy}(t)/J^{\rm e}_{yx}(t)]$ is the coarse entropy increment of a duration-$t$ excursion between $x$ and $y$.
Its $t$ dependence distinguishes route families that share endpoints and integrated transition probabilities.
At equilibrium detailed balance pairs direct fluxes and same-duration excursions pointwise, so both terms in Eq.~\eqref{eq:erw} vanish before integration.
Away from equilibrium, different duration ranges can carry different coarse affinities because they sample different microscopic route families.
This is the information retained by the full waiting-time shape and connects naturally to short-time path inference.

\paragraph{Discussion.---}
Equations~\eqref{eq:uppergap} and \eqref{eq:lowergap} separate two physically distinct forms of ignorance.
The first is microscopic ignorance inside the hidden interior after the boundary path is known.
The second is temporal coarse graining that erases boundary motion between detected events.
ERW removes the second loss whenever the pairwise WTD matrix identifies $G(t)$, while leaving the first loss untouched unless the chosen boundary becomes thermodynamically sufficient for the microscopic likelihood ratio.
This separation is useful experimentally because adding transition detectors can attack the two losses in different ways.
Adding detectors so that source--target closure is retained and the boundary is enlarged refines $Z$ and can directly reduce the upper conditional KL rate.
Adding a redundant event between already resolved boundary states can leave the ideal coarse path unchanged while improving the statistical conditioning of its reconstruction.
The distinction between information gain and estimation gain is therefore visible at the level of detector placement itself.

The full time dependence also gives waiting-time shape a direct physical role.
Short-time powers identify the shortest compatible routes between boundary anchors, with coefficients weighted by products of rates. Intermediate times mix competing routes and repeated visits, whereas long times probe the slow modes of the killed dynamics.
The matrix $\widehat G(s)$ collects these scales into a boundary response, while Eq.~\eqref{eq:schur} extracts the part that must traverse the hidden interior.
The poles and short-time coefficients of different $K_{m\leftarrow\ell}(t)$ must be compatible with a common propagator. Some modes may be absent from individual entries because their residues vanish. Conversely, incompatible reconstructions can reveal missed event channels, nonstationarity, or a breakdown of the finite-state Markov description.
Recent work on blurred transitions, finite temporal resolution, continuous-state extensions, and broader counting statistics provides natural tools for that next step \cite{Ertel2022,Degunther2024,Fritz2025,FritzSeifert2026,MaierFritz2026,Garilli2025,Garilli2026,Raux2026,Buschmann2026,Fiusa2026,Fiusa2026Letter,Seifert2026Review}.

The relation to current-fluctuation bounds remains complementary.
Thermodynamic uncertainty relations compress trajectory information into currents and covariances, whereas the present construction resolves a matrix of first-passage transfer functions \cite{Barato2015,Gingrich2016,Horowitz2020,Ertel2022}.
A recent four-state example shows that an ordinary TUR can slightly exceed the standard single-edge WTD estimator \cite{Gu2026}.
Whether an ordering exists between ERW and TUR remains open.

Waiting between visible events is therefore not an empty interval.
It is a dynamical probe launched from a known post-event state and read out at the source of the next detector firing.
With several directed detectors, cross-waiting-time laws knit these probes into a boundary propagator and an excursion-resolved thermodynamics.
Visible reverse transitions remain useful because they make event-space time reversal directly observable, yet they are not required for the underlying boundary reconstruction.
That distinction separates the experimental information needed to reconstruct hidden motion from the additional symmetry needed to interpret the coarser event process as the standard WTD entropy estimator.

\vspace{1em}

\begin{acknowledgments}
GPT-5.6 Sol assisted with formulating certain proofs and language editing.
The author has independently verified the results.
\end{acknowledgments}

\bibliography{references}

\section{End Matter}

\paragraph{First-passage reconstruction under source--target closure.}
For a detected transition $\ell:u_\ell\to v_\ell$, write $R_\ell=k_\ell|v_\ell\rangle\langle u_\ell|$.
Removing the detected off-diagonal jumps while retaining the original diagonal escape rates gives
\begin{equation}
S=W-\sum_{\ell\in\Lset}R_\ell.
\end{equation}
After event $\ell$ the state is known to be $v_\ell$. Survival without another detected event for time $t$ is propagated by $e^{St}$.
Thus reaching $u_m$ at time $t$ and then making the detected jump $m$ gives
\begin{equation}
K_{m\leftarrow\ell}(t)=k_m[e^{St}]_{u_m v_\ell},
\end{equation}
which is Eq.~\eqref{eq:kernelprop} \cite{Gillespie1977,Harunari2022,VanDerMeer2022}.
Let $\Bset=\Uset=\Vset$, let $P_{\Bset}$ denote the coordinate projector onto $\Bset$, and set $G(t)=P_{\Bset}e^{St}P_{\Bset}$.
For every $x\in\Bset$, closure provides an incoming detected event $\ell_x$ with $v_{\ell_x}=x$ and an outgoing detected event $m_x$ with $u_{m_x}=x$.
Setting $t=0^+$ and then choosing $m=m_x$, $\ell=\ell_y$ gives
\begin{equation}
k_m=K_{m\leftarrow\ell_{u_m}}(0^+),\qquad
G_{xy}(t)=\frac{K_{m_x\leftarrow\ell_y}(t)}{k_{m_x}},
\end{equation}
which proves Eq.~\eqref{eq:reconstruct}.
The event rate obeys $\nu_m=p^{\rm ss}_{u_m}k_m$, hence $p_x^{\rm ss}=\nu_{m_x}/k_{m_x}$.
Finally $G'(0)=P_{\Bset}SP_{\Bset}$, so for $x\neq y$
\begin{equation}
W_{xy}=G'_{xy}(0)+\sum_{\ell\in\Lset:\,u_\ell=y,\,v_\ell=x}k_\ell,
\label{eq:restore}
\end{equation}
while $W_{xx}=G'_{xx}(0)$.

\paragraph{Hidden-excursion kernel.}
Order the states as $\Bset\cup\Hset$ and write
\begin{equation}
S=\begin{pmatrix}A&U\\V&Q\end{pmatrix},\qquad A=P_{\Bset}SP_{\Bset}=G'(0).
\end{equation}
Since a finite irreducible chain reaches the nonempty boundary and eventually fires a visible event, both $Q$ (when present) and $S$ are transient. For real $s\geq0$, $\widehat G(s)=P_{\Bset}(sI-S)^{-1}P_{\Bset}$.
Block inversion of $sI-S$ gives $\widehat G(s)=[sI_{\Bset}-A-U(sI_{\Hset}-Q)^{-1}V]^{-1}$.
A hidden excursion from boundary state $y$ to $x$ has rate density $F_{xy}(t)=[Ue^{Qt}V]_{xy}$, so $\widehat F(s)=U(sI_{\Hset}-Q)^{-1}V$.
Combining these identities yields
\begin{equation}
\widehat F(s)=sI_{\Bset}-G'(0)-\widehat G(s)^{-1},
\end{equation}
which is Eq.~\eqref{eq:schur}.

\paragraph{Excursion-resolved entropy-production rate.}
Define the coarse path by
\begin{equation}
Z_t=\begin{cases}X_t,&X_t\in\Bset,\\ H,&X_t\in\Hset.\end{cases}
\end{equation}
Between consecutive boundary visits, $Z$ records either a direct jump $y\to x$ with rate $\kappa_{xy}=W_{xy}$ or a hidden excursion $y\to H\to x$ of duration $t$ with rate density $F_{xy}(t)$.
Their stationary intensities are
\begin{equation}
J^{\rm d}_{xy}=p_y^{\rm ss}\kappa_{xy},\qquad
J^{\rm e}_{xy}(t)=p_y^{\rm ss}F_{xy}(t).
\end{equation}
First consider a coarse path on $[0,T]$ whose two endpoints lie in $\Bset$.
Its direct jumps are $d:y_d\to x_d$, and its complete hidden excursions are $e:y_e\to x_e$ of durations $t_e$.
Boundary holding factors are $\exp[W_{xx}\times(\text{total time at }x)]$, which are identical for the path and its reverse. Consequently,
\begin{equation}
\ln\frac{\dd P_Z^T}{\dd P_Z^{R,T}}
=\ln\frac{p^{\rm ss}_{z_0}}{p^{\rm ss}_{z_T}}
+\sum_d\ln\frac{\kappa_{x_dy_d}}{\kappa_{y_dx_d}}
+\sum_e\ln\frac{F_{x_ey_e}(t_e)}{F_{y_ex_e}(t_e)}.
\label{eq:boundaryendpoints}
\end{equation}
For a general stationary observation window, an initial or final hidden sojourn can be censored.
It is then incorrect to use only $\ln(p_{z_0}^{\rm ss}/p_{z_T}^{\rm ss})$ as the endpoint contribution. The correct form is
\begin{equation}
\begin{gathered}
\begin{aligned}
\ln\frac{\dd P_Z^T}{\dd P_Z^{R,T}}
={}&\sum_d\ln\frac{\kappa_{x_dy_d}}{\kappa_{y_dx_d}}\\
&+\sum_{e\ {\rm complete}}\ln\frac{F_{x_ey_e}(t_e)}{F_{y_ex_e}(t_e)}+R_T,
\end{aligned}\\[-2pt]
\frac{\mathbb E|R_T|}{T}\longrightarrow0.
\end{gathered}
\label{eq:censoring}
\end{equation}
For example, a path starting in $H$ and first reaching $x\in\Bset$ after time $a$ has an initial factor $[Ue^{Qa}p_{\Hset}^{\rm ss}]_x$. A final uncompleted excursion from $y$ with elapsed time $b$ has factor $\boldsymbol 1_{\Hset}^{\mathsf T}e^{Qb}V_{:y}$. Here $\boldsymbol 1_{\Hset}$ denotes the all-ones vector on $\Hset$, and $V_{:y}$ denotes the $y$th column of $V$.
These factors, and their reversed counterparts, are included in $R_T$.
A path remaining in $H$ for the entire window is invariant as a coarse path under reversal.
For a fixed finite dynamically reversible model, hidden first-passage times and stationary residual times have exponentially decaying tails. The endpoint log factors have finite mean and contribute no asymptotic rate.
Taking the stationary mean gives $\dot D(P_Z\Vert P_Z^R)=\sum_{x\neq y}J^{\rm d}_{xy}\ln(\kappa_{xy}/\kappa_{yx})+\sum_{x,y}\int_0^\infty J^{\rm e}_{xy}(t)\ln[F_{xy}(t)/F_{yx}(t)]\,\dd t$.
Writing $\overline F=\int_0^\infty F(t)\dd t=-UQ^{-1}V$ when $\Hset\neq\varnothing$, elimination of the stationary hidden equation gives
\begin{equation}
(W_{\Bset\Bset}+\overline F)p_{\Bset}^{\rm ss}=0.
\label{eq:boundarybalance}
\end{equation}
The effective matrix $W_{\Bset\Bset}+\overline F$ has zero column sums.
Thus the total incoming and outgoing boundary-changing intensities balance at each boundary state, including both direct and excursion contributions. In particular, the additional terms $\ln(p_y^{\rm ss}/p_x^{\rm ss})$ cancel when the rate ratios are converted to flux ratios.
Pairing $x\to y$ with $y\to x$ then gives
\begin{equation}
\begin{aligned}
\dot D(P_Z\Vert P_Z^R)={}&
\sum_{x<y}(J^{\rm d}_{xy}-J^{\rm d}_{yx})
\ln\frac{J^{\rm d}_{xy}}{J^{\rm d}_{yx}}\\
&+\sum_{x<y}\int_0^\infty
[J^{\rm e}_{xy}(t)-J^{\rm e}_{yx}(t)]
\ln\frac{J^{\rm e}_{xy}(t)}{J^{\rm e}_{yx}(t)}\,\dd t,
\end{aligned}
\end{equation}
which is Eq.~\eqref{eq:erw}, so $\sigma_{\ERW}=\dot D(P_Z\Vert P_Z^R)$.

\paragraph{Universal bound and upper gap.}
Let $P_X^T$ and $P_X^{R,T}$ be the stationary microscopic path measure and its physical time reverse.
For a stationary Markov jump process,
\begin{equation}
\lim_{T\to\infty}\frac{1}{T}D_{\rm KL}(P_X^T\Vert P_X^{R,T})=\sigma.
\end{equation}
The map $X\mapsto Z$ is pointwise in time and commutes with reversal, so data processing gives
\begin{equation}
D_{\rm KL}(P_X^T\Vert P_X^{R,T})\geq D_{\rm KL}(P_Z^T\Vert P_Z^{R,T}).
\end{equation}
Dividing by $T$ and taking $T\to\infty$ yields $\sigma\geq\sigma_{\ERW}$, Eq.~\eqref{eq:directedbound}.
For the same deterministic map and finite path-space relative entropies, the finite-time chain rule is
\begin{equation}
\begin{aligned}
&D_{\rm KL}(P_X^T\Vert P_X^{R,T})\\
&\quad=D_{\rm KL}(P_Z^T\Vert P_Z^{R,T})
+\mathbb E_{z\sim P_Z^T}D_{\rm KL}\!\Biggl[
\begin{aligned}[t]
&P_X^T(\cdot\mid Z=z)\\[-2pt]
&\Vert P_X^{R,T}(\cdot\mid Z=z)
\end{aligned}
\Biggr].
\end{aligned}
\end{equation}
Dividing by $T$ and taking $T\to\infty$ gives Eq.~\eqref{eq:uppergap}.

\paragraph{WTD rate and reverse-closed hierarchy.}
Assume $\Lset=\bar{\Lset}$ and obtain the event-time path $\Gamma$ from $Z$ by retaining only the detected transition labels and the intervals between successive detected events.
The map $Z\mapsto\Gamma$ commutes with physical time reversal.
For $\gamma:\ell_0\xrightarrow{\tau_0}\ell_1\cdots\xrightarrow{\tau_{N-1}}\ell_N$, the forward likelihood contains $\prod_iK_{\ell_{i+1}\leftarrow\ell_i}(\tau_i)$ and the reversed likelihood contains $\prod_iK_{\bar\ell_i\leftarrow\bar\ell_{i+1}}(\tau_i)$. Initial-label and censoring factors have bounded mean contributions as $T\to\infty$.
Since intervals initiated by $\ell$ and followed by $m$ after a delay in $[t,t+\dd t]$ occur with stationary intensity $\nu_\ell K_{m\leftarrow\ell}(t)\dd t$, averaging the log-likelihood ratio gives
\begin{equation}
\dot D(P_\Gamma\Vert P_\Gamma^R)=\sum_{\ell,m}\nu_\ell\int_0^\infty K_{m\leftarrow\ell}(t)\ln\frac{K_{m\leftarrow\ell}(t)}{K_{\bar\ell\leftarrow\bar m}(t)}\,\dd t,
\end{equation}
which is Eq.~\eqref{eq:wtd}, so $\sigma_{\WTD}=\dot D(P_\Gamma\Vert P_\Gamma^R)$.
Data processing for $Z\mapsto\Gamma$, together with the previous contraction $X\mapsto Z$, gives $\sigma\geq\sigma_{\ERW}\geq\sigma_{\WTD}$, Eq.~\eqref{eq:hierarchy}.
The finite-time chain rule for $Z\mapsto\Gamma$ is
\begin{equation}
\begin{aligned}
&D_{\rm KL}(P_Z^T\Vert P_Z^{R,T})\\
&\quad=D_{\rm KL}(P_\Gamma^T\Vert P_\Gamma^{R,T}) +\mathbb E_{\gamma\sim P_\Gamma^T}D_{\rm KL}\!\Biggl[
\begin{aligned}[t]
&P_Z^T(\cdot\mid\Gamma=\gamma)\\[-2pt]
&\Vert P_Z^{R,T}(\cdot\mid\Gamma=\gamma)
\end{aligned}
\Biggr].
\end{aligned}
\end{equation}
Dividing by $T$ and taking the stationary long-time limit yields Eq.~\eqref{eq:lowergap}.

\paragraph{Nested boundaries and full reconstruction.}
For $\Bset_1\subseteq\Bset_2$, obtain $Z^{(1)}$ from $Z^{(2)}$ by retaining states in $\Bset_1$ and mapping states in $\Bset_2\setminus\Bset_1$ together with the hidden symbol of $Z^{(2)}$ to the hidden symbol of $Z^{(1)}$.
This deterministic pointwise map commutes with reversal, so data processing gives
\begin{equation}
\sigma_{\ERW}(\Bset_1)\leq\sigma_{\ERW}(\Bset_2)\leq\sigma,
\end{equation}
which is Eq.~\eqref{eq:boundaryhierarchy}.
If $\Bset=\Omega$, then $\Hset=\varnothing$, $G(t)=e^{St}$ and $S=G'(0)$.
Restoring the detected jumps gives
\begin{equation}
W=G'(0)+\sum_{\ell\in\Lset}k_\ell|v_\ell\rangle\langle u_\ell|.
\end{equation}
Since $X\mapsto Z$ is then the identity, $D_{\rm KL}(P_X^T\Vert P_X^{R,T})=D_{\rm KL}(P_Z^T\Vert P_Z^{R,T})$ for every $T$, and therefore $\sigma_{\ERW}=\sigma$, proving Eq.~\eqref{eq:fullreconstruction}.

\paragraph{One hidden state and extended-real conventions.}
If $\Hset=\{h\}$, the mapping $h\mapsto H$ is a relabeling, not a loss of microscopic information.
With $\lambda=-W_{hh}>0$,
\begin{equation}
F_{xy}(t)=W_{xh}W_{hy}e^{-\lambda t}.
\end{equation}
Let $C_{xy}=p_y^{\rm ss}W_{xh}W_{hy}$.
The excursion contribution for $x<y$ is exactly
\begin{equation}
\int_0^\infty [J^{\rm e}_{xy}(t)-J^{\rm e}_{yx}(t)]
\ln\frac{J^{\rm e}_{xy}(t)}{J^{\rm e}_{yx}(t)}\dd t
=\frac{C_{xy}-C_{yx}}{\lambda}\ln\frac{C_{xy}}{C_{yx}}.
\label{eq:singletonintegral}
\end{equation}
Together with the direct term this equals $\sigma$, also directly by the invertibility of the path map.
For a pair with both fluxes zero, the contribution is zero. One positive flux with a structurally zero reverse gives $+\infty$.
If genuinely one-way microscopic transitions are permitted, the reconstruction formulas and non-subtracted data-processing inequalities remain valid in this extended-real sense.
The gap formulas are asserted only in the finite-entropy regime, to avoid undefined differences such as $\infty-\infty$.

\paragraph{Parameters of numerical examples.}
All numerical data are generated from fully connected six-state Markov networks. Write $k_{i\to j}=W_{ji}$ for the rate from state $i$ to state $j$. For each unordered pair $i<j$, the two directional rates are parameterized as
\begin{equation*}
k_{i\to j}=b_{ij}e^{a_{ij}/2},\qquad
k_{j\to i}=b_{ij}e^{-a_{ij}/2},
\end{equation*}
so that $b_{ij}$ sets the kinetic scale and $a_{ij}=\ln(k_{i\to j}/k_{j\to i})$ is the edge affinity. All rates are expressed in a common inverse-time unit, and the diagonal entries of $W$ are fixed by probability conservation.

For Fig.~\ref{fig:reconstruction}(a), the baseline generator at $f=0$ is obtained by drawing the 15 edge parameters independently as $\ln b_{ij}\sim\mathrm{Unif}[\ln(0.32),\ln(2.4)]$ and $a_{ij}\sim\mathcal{N}(0,0.75^2)$, using a  NumPy \texttt{default\_rng} with seed $20260923$. The detected reversible pairs are $1\leftrightarrow2$ and $3\leftrightarrow4$. The entirely undetected cycle is $1\to5\to6\to1$. Starting from the baseline rates $k^{(0)}$, each forward rate on this cycle is changed to $k^{(0)}e^{f/6}$ and each reverse rate to $k^{(0)}e^{-f/6}$, while every other off-diagonal rate is held fixed. Thus $f$ adds exactly $f$ to the affinity of that hidden cycle. The displayed scan uses $f\in[-2,4]$.

Figures~\ref{fig:reconstruction}(b) and \ref{fig:directed}(b) use the same ensemble of ninety fully connected six-state generators. For every realization and every unordered pair, the parameters are drawn independently as $\ln b_{ij}\sim\mathrm{Unif}[\ln(0.22),\ln(3.8)]$ and $a_{ij}\sim\mathcal{N}(0,1.05^2)$, using a separate fresh NumPy \texttt{default\_rng} with seed $20260923$. In Fig.~\ref{fig:reconstruction}(b), the one-, two-, and three-pair observation sets are respectively $\{1\leftrightarrow2\}$, $\{1\leftrightarrow2,3\leftrightarrow4\}$, and $\{1\leftrightarrow2,3\leftrightarrow4,5\leftrightarrow6\}$. In Fig.~\ref{fig:directed}(b), the observed set for cycle length $q$ is the one-way cycle $1\to2\to\cdots\to q\to1$, for $q=3,4,5,6$.


\end{document}